\documentclass[prl,twocolumn,showpacs,preprintnumbers,superscriptaddress]{revtex4}

\usepackage{times}
\usepackage{subfig}
\usepackage{bm}
\usepackage{graphicx}
\usepackage{amsbsy}
\usepackage{amsmath}
\usepackage{amsfonts}
\usepackage{amsthm}
\usepackage{float}
\usepackage{color}

\begin{document}
 \theoremstyle{plain}
\newtheorem{theorem}{Theorem}
\newtheorem{lemma}[theorem]{Lemma}
\newtheorem{corollary}[theorem]{Corollary}
\newtheorem{proposition}[theorem]{Proposition}
\newtheorem{conjecture}[theorem]{Conjecture}

\theoremstyle{definition}
\newtheorem{definition}[theorem]{Definition}

\title{Quantum Dissension: Generalizing Quantum Discord for Three-Qubit States}
\author {Indranil Chakrabarty}
\author {Pankaj Agrawal}
\affiliation{Institute of Physics, Sainik School Post, 
Bhubaneswar-751005, Orissa, India } 
\author {Arun K Pati}
\email{indranil@iopb.res.in, agrawal@iopb.res.in, akpati@hri.res.in}
\affiliation{Harish Chandra Research Institute, Chhatnag Road, Jhunsi, 
Allahabad-211019, UP, India}

\begin{abstract}
We introduce the notion of quantum dissension for a three-qubit system as a 
measure of quantum correlations. We use three classically equivalent expressions of 
three-variable mutual information. Their differences are zero classically 
but not so in quantum domain. It generalizes the notion of quantum discord 
to a multipartite system. There can be multiple definitions 
of  the dissension depending on the nature of projective measurements done 
on the subsystems. As an illustration, we explore the consequences of these 
multiple definitions and compare them for three-qubit pure and mixed GHZ 
and W states. 
We find that unlike discord, dissension can be negative. This is because
measurement on a subsystem may enhance the correlations in the rest of the 
system. Furthermore, when we consider a bipartite split of the system,
the dissension reduces to discord. This approach can pave a way to 
generalize the notion of quantum correlations in the multiparticle setting.
\end{abstract}


\maketitle
\section{1. Introduction}
Quantum entanglement  plays an important role in the quantum communication
protocols like teleportation \cite{ben,hor}, superdense coding \cite{ben1},
remote state preparation \cite{pati}, cryptography \cite{gisin} and many more.
However, the precise role of entanglement in quantum information processing 
still remains
an open question. It is not clear, whether all the information processing tasks
can be done more efficiently with a quantum system that requires entanglement as
a resource. In addition, precise nature of the quantum correlations is not
well understood  for two-qubit mixed states and multipartite states. It has
been suggested that the quantum correlations go beyond the simple
idea of entanglement. The idea of quantum discord \cite{oll,hen,luo} is to
 quantify all types of quantum correlations including entanglement. It must
 be emphasized here that discord actually supplements the measure of 
entanglement that can be defined on the system of interest. Other measures
 of quantum correlations that have been proposed in the literature 
similar to discord are 
quantum deficit \cite{raja,hor1,dev}, quantumness of correlations \cite{ur}
and quantum dissonance \cite{mw10}.\\

The idea of discord \cite{oll} uses the generalization of two-variable mutual
information to quantum domain. The difference of two classically equivalent expressions
when generalized to quantum setting gives the discord. It was shown
that the discord reduces to von Neumann entropy for a pure bipartite state.
Furthermore, it was found to be nonzero for some of the separable states.
A number of authors have computed
the discord in diverse situations \cite{discord1}. 
In a different approach, entropic methods have been proposed to understand the 
separability and correlations of a composite
state  \cite{raja}. In particular, `quantum deficit' was introduced 
as a measure of quantumness over classical 
correlations. Quantum deficit is the difference of the von Neumann entropy 
of the system and that of the decohered state.
In another approach, Horodecki {\it et al} \cite{hor1} have investigated 
the relation between \textit{local} and \textit{nonlocal}  information  by 
investigating a situation where parties sharing a multipartite state 
distill local information. The amount of information that is lost due to 
the use of classical communication channel is called \textit{deficit}. 
It was shown that the upper bound 
for the deficit is given by the relative entropy distance to so-called 
pseudo classically correlated states. On  the other hand the lower bound is 
the relative entropy of entanglement. 
It was argued  that it was the basic reason for any entangled state to be 
viewed as informationally nonlocal. 
In another piece of work \cite{dev}, a simple information-theoretical measure of the 
one-way distillable local purity was introduced. Interestingly, the author showed 
that his proposed characterization is closely related to a 
previously known operational measure of classical correlations, the one-way
 distillable common randomness.\\

Recently, another measure has been introduced which is called 
quantumness of correlation\cite{ur}. It has been defined 
for bipartite states by incorporating a specific measurement scheme. 
  It was shown that if one uses the  optimal generalized measurement on one of the
 subsystem it reduces the overall state in its closest separable form. It was
 seen that this measure gives a non zero value for all bipartite entangled 
states and zero for separable states. Not only that, it also serves as an 
upper bound to the relative entropy of entanglement.
In Ref \cite{mw10}, authors have given a unified view of the correlations in 
a given quantum state by classifying it into entanglement, dissonance, and
 classical correlations. They used the concept of relative entropy as a 
distance measure of correlations. It has been claimed that their methods 
completely fit into multipartite systems of arbitrary dimensions. In addition
 they showed that dissonance attains a non zero value in the case of pure 
multipartite states. It has also been suggested \cite{ap}
that one single real number may not be able to capture the multifaceted 
nature of the correlations in the multipartite situation. One may need a
vector quantity (a set of numbers) to characterize the correlations. In addition, the notion of 
maximally entangled state may depend on the task that is to be carried out.

In this paper, we generalize the notion of quantum discord from bipartite to 
tripartite systems. Our approach is based on three-variable mutual 
information and its generalization to quantum setting. Classically, 
three-variable
mutual information characterizes the common information contained in
three classical distributions. Therefore, it appears natural to generalize
it to quantum domain. We introduce three 
expressions of the three-variable mutual information, all of which 
are same classically but  differ when conditional entropies are 
generalized to quantum level.  For a tripartite system,  one can make 
measurement on
one-particle, or on more than one-particle to probe the different aspects of the
quantum correlations. This would lead to multiple quantities that can characterize
the correlations. We call these physical quantities ``Quantum Dissension.'' 
In the case of a tripartite state, we shall have two
quantities that will characterize the correlations together.
One can think of the dissension as the information contained in the 
multiparticle quantum state that cannot be 
extracted by either one-particle or two-particle measurements.
We note, as discussed below, in the case of one-particle measurement, the 
dissension can be negative because a measurement on a subsystem 
can enhance the correlations in the rest of the system. Interestingly,
in the case of two-particle measurements, the dissension is just
the discord of a bipartite split of the system.
 We illustrate the idea of dissension with the help of specific three-qubit
pure and mixed states. The pure states have genuine multipartite correlations. 
These states have non-zero dissension with respect to both one and two-particle
measurements. Interestingly, in the case of mixed GHZ and W states, we have
situations analogous to  the two-qubit Werner state. It is known that Werner
state has non-zero discord even for those values of classical mixing
parameter for which it is is not entangled. Similarly, we find that the mixed 
GHZ and W states have non-zero 
 dissension for all values of classical mixing parameter. 
 Therefore, dissension may be characterizing the quantumness in multiparticle 
system that goes beyond entanglement.
We also provide an example where dissension due to two-particle measurement 
is zero, but it is non-zero when we make only one-particle measurement.

The organization of our paper is as follows. In section 2, we introduce the 
notion of three-variable mutual information and then define the quantum 
dissensions for one-particle and two-particle measurements.  In section 3, we 
calculate the dissension for pure three-qubit states like GHZ and 
W states. In section 4, we illustrate the dissension for generalized Werner 
kind of three-qubit mixed states. We also discuss the case of a biseparable
state. Finally, in section 5, we conclude and mention
future directions of explorations.

\section{2. Quantum Dissension }

In classical information theory \cite{cover}, one can quantify the relationship
between two random variables $X$ and $Y$  by a quantity called mutual information $I(X:Y)=H(X)-H(X|Y)$
where $H(X),H(X|Y)$ are the entropy of $X$ and the conditional entropy of $X$ given that $Y$ has 
already occurred. Mutual information actually gives the measure of the reduction in the uncertainty 
about one random variable because of the occurrence of the other random variable. Since 
$H(X|Y)=H(X,Y)-H(Y)$, so there is another alternative 
expression for the mutual information $J(X:Y)=H(X)+H(Y)-H(X,Y)$. Classically, these two expressions 
of the mutual information are identical. We can generalize these expressions to quantum domain by
substituting random variables $X,Y$ by  the density matrices $\rho_X, \rho_Y$ and the Shannon entropies
$H(X), H(Y)$ by the Von Neumann entropies (e.g: $H(X)=H(\rho_X)=-{\rm Tr}[\rho_X \log(\rho_X)$]). With these
substitutions, one can define $J$ in the quantum case. But to define $I$, one needs to specify
the conditional entropy $H(X|Y)$. To specify this conditional entropy,  
the measurement of Y is defined by a set of one dimensional
projectors $\{\pi^Y_i\}$ \cite{oll}. Here, the subscript $i$ refers to the outcome of this measurement.
This gives us the quantum analogue of $I(X:Y)$ as
\begin{eqnarray}
  I(X:Y)=H(X)-H(X|\{\pi^Y_i\}),\
\end{eqnarray}
where $H(X|\{\pi^Y_i\}) = \sum_j p_jH(\rho_{X|\pi^Y_i})$, 
$\rho_{X|\pi^Y_i}=\frac{\pi^Y_i \rho_{XY} \pi^Y_i}{Tr(\pi^Y_i \rho_{XY})}$  and 
$p_i$ is the probability of obtaining the $i$th outcome. 
It is clearly evident that these two
expressions for $I(X:Y)$ and $J(X:Y)$ are not identical in quantum theory. 
The quantum discord is
the difference \cite{oll,hen,luo}
\begin{eqnarray}
D(X:Y) = J-I = H(Y)-H(X,Y)+H(X|\{\pi^Y_i\}).
\end{eqnarray}
 This is to be minimized over 
all sets of one dimensional projectors $\{\pi^Y_i\}$.

It is possible to generalize the discord to multipartite systems by 
considering multiparticle-measurement.
In this paper, we introduce the notion of Quantum Dissension in the context 
of three-qubits. In the case of
three qubits, there can be two types of projective measurements. These are 
one-particle  projective measurements
and two-particle projective measurements. These measurements can be performed
 on different subsystems. 
This would lead to multiple definitions of the quantum dissension.
We focus on two scenarios. In one case, we consider all possible one-particle 
projective measurements. In the second scenario,
we make all possible two-particle measurements. 

 To obtain the definition of the dissension
 for these two scenarios, we consider 
three-variable classical mutual information \cite{cover}. It is defined as
\begin{eqnarray}
 I(X:Y:Z) = I(X:Y) - I(X,Y|Z).
\end{eqnarray}
Here, $ I(X,Y|Z)$ is the 
conditional mutual information
\begin{eqnarray}
I(X,Y|Z) =  H(X|Z)+H(Y|Z)-H(X,Y|Z).
\end{eqnarray}
  Both $I(X:Y)$ and $I(X,Y|Z)$ are non-negative. However, there may exist a 
situation, when the conditional mutual information is greater than the mutual 
information. It happens when knowing the variable $Z$ enhances the correlation
between $X$ and $Y$. In such a case, the three-variable mutual information is 
negative. One well known example where it occurs is  modulo 2 addition of  
two binary random variables. This is XOR gate. Suppose we add $X$ and $Y$ variables 
and get the variable $Z$. Let the variables $X$ and $Y$ be independent, then $I(X:Y)$
is zero. However, once we know the value of $Z$, knowing the value of $Y$ 
determines the value of $X$ uniquely. So the knowledge of $Z$, enhances the
correlations between $X$ and $Y$. Therefore, $I(X,Y|Z)$ is non-zero. This
implies that $I(X:Y:Z)$ is negative. This negative value captures a certain
aspect of the correlations among the variables $X,Y,$ and $Z$. We discuss 
an example for the quantum case later in this section.

 In the case of three random variables, we can obtain many 
 different classically equivalent expressions
 for $ I(X:Y:Z)$. To achieve our goal, we obtain three different
 equivalent expressions. In one of the cases, the expression
 does not have any conditional entropy, while in the
 other two cases, the expression has conditional entropies
 with respect to one variable only or two variables only. 
 In this way, after generalization to quantum domain, we can explore 
the quantum correlations
of different partitions of the quantum system. Using the chain rule in 
the definition (4), we obtain
\begin{eqnarray}
I(X:Y:Z)= H(X,Y)-H(Y|X)-H(X|Y)\nonumber\\- H(X|Z)-H(Y|Z)+H(X,Y|Z). 
\end{eqnarray}
We can convert the above expression that involves conditional entropies to 
that containing only entropies and joint entropies. We obtain
\begin{eqnarray}
J(X:Y:Z)= [H(X)+H(Y)+H(Z)] \nonumber\\- [H(X,Y) + H(X,Z) +H(Y,Z)]+H(X,Y,Z).
\end{eqnarray}
Using the chain rule $H(X,Y,Z) = H(Y,Z) + H(X|Y,Z)$, we can define the 
three-variable mutual information involving two-variable conditional 
entropies. This gives another equivalent expression
\begin{eqnarray}
 K(X:Y:Z)=  [H(X)+H(Y) +H(Z)]\nonumber\\- [H(X,Y)+H(X,Z)] + H(X|Y,Z).
\end{eqnarray}
These three expressions for the three-variable mutual information are classically equivalent, but not
so in quantum domain. The difference of the three definitions can capture various aspects of the quantum
correlations. In the next subsection, we will generalize these definitions to quantum domain.

In principle, for a given quantum system the amount of information one can extract from it
depends on the nature and choice of the measurement.  However, it has been shown \cite{ad}
that the discord does not changes if we use POVM measurements instead of projective 
measurements. As dissension can be written in terms of the discord, as shown below,
one may expect that the situation will remain same.
Therefore, in the next section we built up 
our definitions on the basis of projective measurement . But one can choose POVM
measurements also.


\subsection{Quantum Dissension for One-Particle Projective Measurement}
Here, we extend the definitions of the three-variable mutual information 
$I(X:Y:Z)$ and $J(X:Y:Z)$ to the quantum domain. Let us consider a three-qubit state $\rho_{XYZ}$, 
where $X,Y,Z$ refer to the first, second and the third qubit. The extension of the definition of 
$J(X:Y:Z)$ is straightforward. It is obtained by replacing the random variables 
by the density matrices and the Shannon entropies by the Von Neumann entropies. 
The extension of  the expression for $I(X:Y:Z)$
requires appropriate extension of the conditional entropies.
It is given by
\begin{eqnarray}
 I(X:Y:Z)=H(X,Y)-H(Y|\{\pi_j^X\})- H(X|\{\pi_j^Y\})\nonumber\\
-H(X|\{\pi_j^Z\})-H(Y|\{\pi_j^Z\})+H(X,Y|\{\pi_j^Z\}),
\end{eqnarray}
where $H(X|\{\pi^Y_j\})=\sum_j p_jH(\rho_{X|\pi^Y_j})$,
$\rho_{X|\pi^Y_j}=\frac{\pi^Y_j \rho_{XY} \pi^Y_j}{Tr(\pi^Y_j \rho_{XY})}$ and
 $p_j$ is the probability of obtaining the $j$th outcome. 
Here,  $H(X|\{\pi^Y_j\})$ is the average Von Neumann entropy of the qubit $X$,
 when the projective measurement is done 
on the subsystem $Y$ in the general basis $\{|u_1\rangle = 
\cos(t)|0\rangle+\sin(t)|1\rangle, |u_2\rangle = \sin(t)|0\rangle- 
\cos(t)|1\rangle\}$ 
(where $t\in [0,2\pi$] ). Similarly, one can write down the equivalent 
expressions for $H(X|\{\pi_j^Z\}),H(Y|\{\pi_j^Z\}),H(Y|\{\pi_j^X\})$.
Furthermore, $H(X,Y|\{\pi_j^Z\})=\sum_j p_jH(\rho_{X,Y|\pi^Z_j})$, 
$\rho_{X,Y|\pi^Z_j}=\frac{\pi^Z_j \rho_{XYZ} \pi^Z_j}{Tr(\pi^Y_j \rho_{XYZ})}$. 
It is the average Von Neumann entropy of the subsystem '$XY$', 
when the projective measurement is carried out on the qubit $Z$. 
 $H(X,Y)$ refers to Von Neumann entropy of the density matrix $\rho_{X,Y}$.  
To define dissension for the single particle 
projective measurement, we consider the difference between  
$I(X:Y:Z)$ and $J(X:Y:Z)$. This difference is given by

\begin{eqnarray}
 D_1(X:Y:Z)=  I(X:Y:Z)-J(X:Y:Z) \nonumber \\= H(X,Y|\{\pi_j^Z\})+[H(X,Z)+H(Y,Z)\nonumber\\+
  2H(X,Y)]-H(X,Y,Z)- [H(X|\{\pi_j^Y\})\nonumber\\ + H(X|\{\pi_j^Z\})+H(Y|\{\pi_j^Z\})+H(Y|\{\pi_j^X\})]\nonumber \\
 - [H(X)+H(Y)+H(Z)].
\end{eqnarray}

One can minimize this over all possible one-particle measurement projectors.
So mathematically the expression for the dissension is given by, 
$\delta_1= {\rm min}(D_1(X:Y:Z))$. For single-particle projective measurements,  the above 
expression is the most general one in the sense that it includes all possible 
one-particle projective measurements.  As a consequence of which the dissension
$ \delta_1$, may reveal the maximum possible quantum correlations. We note that dissension is not symmetric
with respect to the permutations of the subsystems $X,Y$ and $Z$, as in the case of discord. \\

\textbf{{Lemma 1}}: For an arbitrary  pure three-qubit state $J(X:Y:Z) = 0$. Therefore, $D_1 =  I(X:Y:Z)$.\\

\textbf{{Proof}}: For a pure three-qubit state $H(X,Y,Z) = 0$, This is because $\rho_{XYZ}$ being 
a pure state has no uncertainty. Furthermore, Von Neumann entropies for the subsystems are related
as  $H(X) = H(Y,Z),~~ 
H(Y) = H(X,Z)$, and $H(Z) = H(X,Y)$. Therefore,
$J(X:Y:Z) = 0$ and $D_1 =  I(X:Y:Z)$ for a pure three-qubit state.\\

\textbf{{Lemma 2}}: For an arbitrary pure three-qubit system, 
$H(X,Y|Z) = H(X,Z|Y) = H(Y,Z|X) = 0$.\\

\textbf{{Proof}}: In the quantum domain, $H(X,Y|Z) = H(X,Y|\{\pi_j^Z\})= 
\sum_j p_jH(\rho_{X,Y|\pi^Z_j})$. Here, $\rho_{X,Y|\pi^Z_j}$ is
the density matrix of the system after the measurement has been performed 
on the subsystem $Z$. After the measurement,
for each projector, the state of the subsystem XY is a pure state. Therefore, 
the Von Neumann entropy 
$H(X,Y|Z) = H(\rho_{X,Y|\pi^Z_j}) = 0$. Similarly,
von Neumann entropies $H(X,Z|Y)$ and $H(Y,Z|X)$ are also zero.
 
\subsection{Quantum Dissension for Two-Particle Projective Measurement}
For a three-qubit system, we can also make measurement on a two-qubit 
subsystem. This will probe different aspects
of the quantum correlations.  We also define dissension involving all two-qubit measurements. For this
purpose, we define the quantum analogue of the classical mutual information 
$K(X:Y:Z)$ as follows.

\begin{eqnarray}
 K(X:Y:Z)= [H(X)+H(Y)+H(Z)]\nonumber\\ - [H(X,Y)+H(X,Z)]+H(X|\{\pi^{Y,Z}_j\})
\end{eqnarray}
 where $H(X|\{\pi^{YZ}_j\})=\sum_j p_jH(\rho_{X|\pi^{YZ}_j})$;
$\rho_{X|\pi^{YZ}_j}=\frac{\pi^{YZ}_j \rho_{XYZ} \pi^{YZ}_j}{Tr(\pi^{YZ}_j 
\rho_{XYZ})}$. It  is the average Von Neumann entropy of the qubit $X$,
when the projective measurement is carried out on the subsystem '$YZ$' 
in the general basis 
$\{ |v_1\rangle = \cos(t)|00\rangle+\sin(t)|11\rangle, 
|v_2\rangle = -\sin(t) |00\rangle+\cos(t)|11\rangle,|v_3\rangle = 
\cos(t)|01\rangle+\sin(t)|10\rangle,|v_4\rangle =  -\sin(t)|01\rangle+\cos(t) 
|10\rangle\}$ and $p_j$ is the probability of obtaining the $j$th outcome. 
Here, $H(X),H(Y),H(Z),H(X,Y),H(X,Z)$ 
represents the Von Neumann entropies of the subsystems, 
$\rho_X,\rho_Y,\rho_Z,\rho_{XY},\rho_{XZ}$, respectively.
To define the dissension, we take the difference

\begin{eqnarray}
 D_2(X:Y:Z) =  K(X:Y:Z) - J(X:Y:Z)\nonumber \\ 
=  H(X|\{\pi^{YZ}_j\})+H(Y,Z)-H(X,Y,Z).
\end{eqnarray} 
  Like one-particle projective measurement case, here also we define dissension 
as, $\delta_2= {\rm min}(D_2(X:Y:Z))$. 
Furthermore, as in the case of discord, this quantity is not symmetric 
under the permutations of $X,Y$ and $Z$. If we choose to make a measurement
on 'XY' subsystem, instead of 'YZ' subsystem, we can obtain the corresponding
expression for $D_2$ by interchanging $X$ and $Z$. Similarly, we can interchange
$X$ and $Y$ if we make a measurement on 'XZ' subsystem. In each case, $D_2$
will reduce to discord with appropriate bipartite split, as discussed below in Lemma
$4$.\\

\textbf{{Lemma 3}}: For an arbitrary pure three-qubit system, $H(X|Y,Z) = H(Y|X,Z) = H(Z|X,Y) = 0$. Therefore, $D_2 = H(X)$ and
the dissension is given by the Von Neumann entropy of the bipartite partition.\\

\textbf{{Proof}}: We have defined:  $H(X|Y,Z) = H(X|\{\pi^{YZ}_j\})=\sum_j p_jH(\rho_{X|\pi^{YZ}_j})$. After the measurement,
the system is in a product state of the state of the $X$ and the projected state of the $YZ$ subsystem, which is a
pure state. Therefore, its Von Neumann entropy is zero, i.e., $H(\rho_{X|\pi^{YZ}_j}) = 0$ for all $j$. It implies that $H(X|Y,Z) =0$.
Similarly, $H(Y|X,Z) = H(Z|X,Y) = 0$. Furthermore, for any pure three-qubit state, $H(X) = H(Y,Z)$ and $H(X,Y,Z) = 0$. Therefore,
$D_2 = H(X)$.

We note that the above Lemma is for three-qubit case. For multi-qubit case 
one can define the dissension along
the same line. In that case the expression will be non-trivial and it is not going to be the Von Neumann
entropy of the bipartite partition \cite{cap}.\\

\textbf{{Lemma 4}}: Dissension is related to discord as : 
\begin{eqnarray}
D_1(X:Y:Z)&=& D(X,Y:Z) - D(X:Z) - D(Y:Z) \nonumber \\ 
     &&          - D(X:Y) - D(Y:X), \\ 
 D_2(X:Y:Z)&=& D(X:Y,Z). 
\end{eqnarray}
Therefore, in the case of a tripartite state, $D_2$ is just discord with a bipartite split of
the system.\\

\textbf{{Proof}}: The proof of this lemma is straightforward. 
To prove the relation $(12)$, we start with the definition of 
 $D_1$ which is given in $(9)$. We now group terms together 
 and rewrite them in terms of discord $D$, using $(2)$.
We notice that the fourth, sixth, ninth, tenth and eleventh terms
in $(9)$ combine to give the last two terms of $(12)$. 
By adding and subtracting $H(Z)$, we can get the remaining three
terms. The relation for $D_2$ is obvious from $(2)$ and $(11)$.

\subsection{Quantum Mutual Information Can be Negative}

  Classical three-variable mutual information is defined in Equations (3) and (4).
  Its generalization to quantum domain has been discussed in this section.
  As discussed earlier, when $I(X:Y)$ is smaller than $I(X,Y|Z)$,
 then $I(X:Y:Z)$ can be negative. It happens when knowing 
 the state $Z$ enhances the quantum correlations between states X and Y. 
 For a simple example, consider pure three-qubit GHZ state. If we trace 
 out one qubit (say $Z$), the reduced density matrix is a
mixture of product states. If we compute mutual information with 
measurement in Hadamard basis, then the mutual information
 $I(X:Y)$ is zero.  However, a measurement in Hadamard  basis on
 the qubit $Z$ reduces the subsystem of the qubits $X$ and $Y$ to 
 a Bell State. One can compute that $I(X,Y|Z)=2$. As a result 
 $I(X:Y:Z)=-2$. Here, we see that a measurement in appropriate
basis on a subsystem can enhance the correlation in the rest of the system. 
So the conditional
mutual information can be larger than the mutual information, thus making 
three-variable
mutual information negative. This should not be regarded as a drawback of the 
definition. On the contrary this may give some new insight into the true 
nature of quantum correlations in multipartite setting. We may also note 
that different measures of correlations for multipartite situations 
may capture different aspects of the quantumness. 

\subsection{Comparison with other measures for three-qubit states}

In reference \cite{mw10}, the authors gave an unified view of total correlation for multi party states in terms of relative entropy. They defined entanglement ($E$), discord ($D$) and  dissonance ($Q$) for a given state $\rho$ as  
$E=min_{\sigma \in {\cal S}} S(\rho || \sigma), D=min_{\xi \in {\cal C}}S (\rho || \xi), Q=min_{\xi \in {\cal C}}S (\sigma || \xi)$. Here $S(. ||. )$ represents the relative entropy. The states $\sigma$ and $\xi$ are the closest states to the the state $\rho$ in the sets of  separable states ($\cal S$) and  classical states ($\cal C$) respectively.
However, our approach is based on three-variable mutual information, not on relative entropy. 
We are looking at the information contained in a specific state, while in the case of relative entropy, 
the distances from separable or classical states are used.  Our philosophy is that one number is not enough
to characterize the quantum correlations of a multipartite state. Therefore, we consider two separate types of
measurements. Depending on what
information processing task one wishes to carry out, one set of measures may be more useful than the others.

\section{3. Quantum Dissension for Pure Three-Qubit states} 

In this section we present quantum dissension for pure three-qubit states. 
As mentioned earlier, one can carry out both the single-particle and two-particle 
 projective measurements in the most general basis. We illustrate 
the usefulness of these definitions by considering pure three-qubit GHZ and 
W states.      


\subsection{Quantum Dissension for GHZ state}

Let us consider a pure three-qubit GHZ state
\begin{eqnarray}
|GHZ\rangle_{ABC}=\frac{1}{\sqrt{2}} ( |000\rangle+|111\rangle ).
\end{eqnarray}
First, we calculate the dissension when the projective measurement is 
carried out on one particle.  After tracing out two qubits, the one-qubit 
density matrices representing the individual subsystems are given by 
$\rho_A=\rho_B=\rho_C = \frac{I}{2}$
with the von Neumann entropies equal to one, i.e., $H(A)=H(B)=H(C)=1$. 
Similarly, by tracing out any one of the three qubits, we obtain the reduced 
density matrices of the subsystems as
\begin{eqnarray}
 \rho_{AB}=\rho_{BC}=\rho_{CA}= \frac{1}{2} (|00\rangle\langle 00|+ 
|11\rangle\langle 11| ).
\end{eqnarray}
Consequently, the entropies of these density matrices are given by 
$H(AB)=H(BC)=H(CA)=1$. Since $|GHZ\rangle$ is a pure state,
so the joint Von Neumannn entropy  $H(ABC)$ vanishes.

Next, we consider the conditional entropy of the two systems, when the 
projective measurement is done on the third system and we find it to be 
zero, i.e., $H(AB|\{\pi_j^C\})=0$ (using Lemma 2).
We also find the conditional entropies of one qubit system when the 
projective measurement is done on any one of the other two-qubit systems. 
These are given by
\begin{eqnarray}
&&H(A|\{\pi_j^B\})=H(A|\{\pi_j^C\})={}\nonumber\\&& H(B|\{\pi_j^C\})=H(B|\{\pi_j^A\})={}\nonumber\\&&-(\frac{1-\cos(2t)}{2})\log_2{\frac{1-\cos(2t)}{2}}{}\nonumber\\&&-(\frac{1+\cos(2t)}{2})\log_2{\frac{1+\cos(2t)}{2}}.
\end{eqnarray}
Consequently, the expression for $D_1$ in case of GHZ state is given by
\begin{eqnarray}
 D_1=1+4[\frac{1-\cos(2t)}{2}\log_2{\frac{1-\cos(2t)}{2}}\nonumber\\ +\frac{1+\cos(2t)}{2}\log_2{\frac{1+\cos(2t)}{2}}].
\end{eqnarray}

We have plotted $D_1$ for the GHZ state for the single-particle projective 
measurement  as a function of $t$ in the Figure 1 $(i)$. 
It is a oscillating function which varies between $[-3,1]$. By minimizing over 
all possible bases, the dissension is given by $\delta_1=-3$. The
part of this calculation was performed using the 
mathematica package QDENSITY \cite{thi}. 
For GHZ state, in case of two-particle projective measurement, the 
conditional entropy is zero and
the dissension reduces to the  bipartite entanglement present in the system 
and is equal to one (see Lemma 3). Therefore dissension for the state is
$(\delta_1, \delta_2) = (-3.00, 1.00)$.

\subsection{Quantum Dissension for W state}

Here, we consider another inequivalent class of pure three-qubit state which 
is known as W state. This is given by
\begin{eqnarray}
|W\rangle_{ABC}=\frac{1}{\sqrt{3}} (|100\rangle+|010\rangle+|001\rangle ).
\end{eqnarray}
The density matrices representing each of the one qubit subsystem of the 
three qubit W state are given by
\begin{eqnarray}
\rho_A = \rho_B = \rho_C = 
\frac{1}{3}[2|0\rangle\langle 0|+|1\rangle\langle 1|].
\end{eqnarray}
 The Von Neumann entropies of these density matrices are found to be 
$H(A)=H(B)=H(C)=0.92$.
Similarly, by tracing out any one qubit we obtain the two qubit density 
matrices representing two-qubit subsystem of the W state. These are given by 
\begin{eqnarray}
\rho_{AB}=\rho_{BC}=\rho_{CA}=\frac{1}{3}[|00\rangle\langle 00|+ 
|01\rangle\langle 01|\nonumber\\+|01\rangle\langle 10| +|10\rangle\langle 01| 
+ |10\rangle\langle 10|].
\end{eqnarray}
Hence, the Von neumann entropies of the two-qubit subsystems are given by
$H(AB)=H(BC)=H(AC)=0.92$. Since the $W$ state is a pure state, the 
joint Von Neumannn entropy for the $W$  state is zero, i.e., $H(ABC)=0$ . 
Then, we 
find out all other conditional entropic quantities required for finding out
 the dissension when the projective measurement is carried out on a single 
qubit system. The conditional entropy of any two-qubit system when the 
projective measurement is carried out on the third qubit is zero, i.e.,
 $H(AB|\{\pi_j^C\})=0$ (see Lemma 2).
In addition, the conditional entropies of one qubit systems when the 
projective measurement is done on any one of the remaining qubits, are
 given by
\begin{eqnarray}
H(A|\{\pi_j^B\})=H(A|\{\pi_j^C\})=H(B|\{\pi_j^C\})=  \nonumber \\
 H(B|\{\pi_j^A\})
=1 + {1 \over 2}\, (p \, H(1+x_{+}, 1-x_{+})  \nonumber \\
            +  (1-p)\,H(1+x_{-}, 1-x_{-}) ),
\end{eqnarray}
where $p = {(3 + \cos(2t))\over 6}$ and,
\begin{eqnarray}
H(x,y) & = & -x\, \log_{2}(x) -y\,\log_{2}(y),
\end{eqnarray}
\begin{eqnarray}
x_{\pm} & = & { \sqrt{(5\pm 3\, \cos(2t))\,(1 \mp \cos(2t))} \over (3 \pm \cos(2t))}.
\end{eqnarray}

On using the above equations, the expression for $D_1$ is given by
\begin{eqnarray}
D_1=H(AB)-4H(A|\{\pi_j^B\}).
\end{eqnarray}

On substituting the values of these entropic quantities, the expression $D_1$ for the W state is plotted in Figure 1$(ii)$. In this case, we see from the plot that the dissension is $-1.74$. The computations here and below were performed using the mathematica package QDENSITY \cite{thi}.

Like the GHZ state, for the two-particle projective measurements, the conditional entropy is zero and
the dissension reduces to the  bipartite entanglement present in the system. For the W state, it is equal to $.92$ (see Lemma 3
). So the dissension for the W state is $(\delta_1, \delta_2) = (-1.74, 0.92)$.

\section{4. Quantum Dissension for Mixed Three-Qubit states}

In this section, we analyze quantum dissension for three-qubit mixed states. 
We obtain $D_1$ and
$D_2$ as a function of the angle $t$ and the classical probability of mixing 
$a$ after carrying out the projective measurement in the most general basis. 
The case for the mixed state is illustrated by
the pseudo-pure GHZ state with $\rho_{\rm GHZ} = 
(1-a)\frac{I}{8}+a|GHZ\rangle\langle GHZ|$ 
and  the pseudo-pure $W$ state with
$\rho_{\rm W} = (1-a)\frac{I}{8}+a|W\rangle\langle W|$. (Here $I$
is the random mixture.)
These mixed states can be viewed as the generalization of the Werner state 
for the case of three-qubit states. Specifically, the mixed GHZ state can be 
thought of as a mixture of a random state and
a "maximally" entangled state. As we shall see, the dissension for this 
state is non zero unless it is a random mixture.

\subsection{Quantum Dissension for mixed GHZ state}

Let us consider a three-qubit mixed GHZ state
\begin{eqnarray}
\rho_{\rm GHZ}=(1-a)\frac{I}{8}+a|GHZ\rangle\langle GHZ|.
\end{eqnarray}
 After tracing out any two qubits, the reduced density matrices representing 
the individual subsystems are given by the random mixture, i.e., 
$\rho_A=\rho_B=\rho_C= \frac{I}{2}$.
The Von Neumann entropies of these subsystems are all equal to one.
Similarly, by tracing out a single qubit one can obtain the two-qubit density 
matrices given by
\begin{eqnarray}
\rho_{AB}=\rho_{BC}=\rho_{CA} & =& 
\frac{1+a}{4}[|00\rangle\langle 00|+|11\rangle\langle 11|]{}\nonumber\\
& +& \frac{1-a}{4}[|01\rangle\langle 01|+|10\rangle\langle 10|]
\end{eqnarray}

We have calculated $D_1$ and $D_2$ in general basis. Since the expression 
for the quantum dissensions for $\rho_{\rm GHZ}$ are quite involved and 
lengthy, we do not provide these analytic expressions in this manuscript.  
We have plotted the $D_1$ and  $D_2$ in Figures 2$(i),(iii)$. Here, we can
see that that for $a=1$, this gives us the back the  $D_1$ of the GHZ state 
which is given in Figure 1$(i)$. Furthermore the dissensions  
$\delta_1$ and  $\delta_2$ are non zero for any non-zero values of $a$. 
This is like the two-qubit Werner state. We also notice that $D_2$ 
 reduces to that of the pure GHZ state for $a = 1$ 
as expected.

\subsection{Quantum Dissension for mixed W state }

Here, we evaluate the quantum dissension for the mixed three-qubit W state. 
The state is given by
\begin{eqnarray}
\rho_{\rm W}=(1-a)\frac{I}{8}+a|W\rangle\langle W| .
\end{eqnarray}

After tracing out any two qubits, the reduced density matrices representing 
the one qubit subsystems are given by
\begin{eqnarray}
\rho_A=\rho_B=\rho_C=\frac{3+a}{6}|0\rangle\langle 0|+ 
\frac{3-a}{6}|1\rangle\langle 1|.
\end{eqnarray}
Similarly, one can trace out a single qubit to obtain the density matrices 
of two-qubit subsystems as
\begin{eqnarray}
&&\rho_{AB}=\rho_{BC}=\rho_{CA}{}\nonumber\\&&=[\frac{3+a}{12}] 
[|00\rangle\langle 00|+|01\rangle\langle 01|+|10\rangle\langle 10|]{}\nonumber\\&&+[\frac{1-a}{4}]|11\rangle\langle 11|+\frac{a}{3}[|01\rangle\langle 10| 
+ |10\rangle\langle 01|].
\end{eqnarray}

We evaluate the Von Neumann entropies of these above density matrices, 
all the conditional entropies (for both the one-particle and the two-particle 
measurement) and the joint entropy of three-qubit Mixed W state.
Here also, we do not provide analytic expressions for dissension for 
$\rho_{\rm W}$ as they are quite long and beyond the scope of the present 
paper. In the Figures 2$(ii)$ and $(iv)$, we show $D_1$ and  $D_2$ 
as a function of the classical probability of mixing $a$ as well as the 
angle $t$. It is evident from the Figure   2 $(ii)$ that for $a=1$, it gives 
us back 
the curve which is obtained in the Figure 1 $(ii)$. 
Similarly, in the Figure  2$(iv)$, $D_2$  reduces
 to that of the pure W state for $a = 1$. 

\subsection{Quantum Dissension for a Biseparable State }

   In this subsection, we give an example of a state for which
   $\delta_1$ is non-zero, but $\delta_2$ is zero. The state is
   \begin{eqnarray}
\rho_{\rm bi}& = & a\,(|000\rangle \langle 000| + |011\rangle \langle 011| + |000\rangle \langle 011| + \nonumber \\
                    & & |011\rangle \langle 000|)+ b \, (|001\rangle \langle 001| + |010\rangle \langle 010| - \nonumber \\
                    & & |001\rangle \langle 010| - |010\rangle \langle 001|),
\end{eqnarray}
where $a + b = {1 \over 2}$.  One can compute the dissension for this state as described above. One would find
  $(\delta_1, \delta_2) = (1.00,0.00)$. One can understand these values by rewriting
  this state as a biseparable state
   
\begin{eqnarray}
\rho_{\rm bi}= 2 a\, |0\rangle |\varphi^+\rangle \langle 0| \langle \varphi^+|   + 2b\, |0\rangle |\psi^-\rangle \langle 0| \langle \psi^-|,
\end{eqnarray}   
 where $ |\varphi^+\rangle=\frac{1}{\sqrt{2}}(|00\rangle+|11\rangle),  |\psi^-\rangle=\frac{1}{\sqrt{2}}(|01\rangle-|10\rangle)$.  
 For this state, $D_2$ is zero since the state is a product state in 
the partition $1:(23)$. However,
 $D_1$ is due to single-particle measurement and it is non-zero. One can compute the concurrence for
 the qubit subsystem $2$ and $3$ and find that this subsystem is indeed 
entangled. Therefore,
 as argued earlier, the correlations of a tripartite state needs to be characterized by both
 $(\delta_1, \delta_2)$. This example also shows that one needs to compute $D_2$ for
 all bipartite partitions.

\section{5. Conclusion}
    
In this paper we have introduced the notion of Quantum Dissension for 
multipartite systems. Starting from three-variable mutual information
definition, we obtained three equivalent expressions. When these expressions are
generalized to quantum domain, then  from the difference
 between two such expressions, we obtain two
kinds of dissensions. Classically, these differences are zero but in 
quantum domain they are not. This generalizes the notion of quantum discord
to tripartite systems. 
We have multiple definitions of the dissension that can capture different aspects of the 
quantum correlations. We have illustrated
the notion of quantum dissension using three-qubit pure and mixed sates. 
Specifically, we have taken pure and mixed GHZ as well as 
W states and calculated the dissensions. We have focused on two 
such definitions of dissensions.
One of the definitions $\delta_{1}$ involves only one-particle measurements, 
while the second definition $\delta_{2}$ involves only two-particle 
measurements. We find that dissension can be negative. It just reflects the 
fact that a  measurement on a subsystem can enhance the correlation of the rest 
of the system. This is a new feature that emerges when we deal with 
multiparticle systems. One can interpret the dissension as the information 
contained in the multiparticle quantum sate that cannot be
extracted by one-particle or two-particle measurements. We have also
obtained the relations between $D_1$, $D_2$ and the discord $D$.
It turns out that $D_2$ is just the discord for 
the bipartite split. However, if we generalize discord to a system of more than three
qubits, then appropriate $D_2$ would not be discord (and even can 
be negative), similar to the quantity like $D_1$ \cite{cap}.
We also note that the dissension is non-zero for all values of the 
classical mixing parameter
for the mixed GHZ and W states. The mixed GHZ state that we have 
considered can be thought of as a generalization of the
two-qubit Werner state. In the case of Werner state, entanglement 
vanishes for certain values of the mixing parameter, but
discord is non-zero. 
We have also discussed one example where the dissension due to one-particle
measurement is non-zero but that due to two-particle measurement is zero.
In future, one may like to put bounds on the values of the dissension for three
qubit system and higher dimensional systems. 
One may also like to relate the notion of dissension to the successes 
of various communication protocols. This may help in providing an operational 
meaning to dissension. Following our approach, one can generalize the 
notion of dissension beyond three-qubit  system \cite{cap}. It will be 
interesting to see if the multipartite dissensions play any role 
in giving extra power to quantum computing in the  absence of the entanglement.

\vskip 3cm

\textit{Acknowledgments:} We thank R. Horodecki for his useful remarks. 
We also acknowledge useful discussions with K. Modi.

\begin{widetext}
\begin{center}
\begin{figure}
\[
\begin{array}{cc}
\includegraphics[height=4.0cm,width=5.5cm]{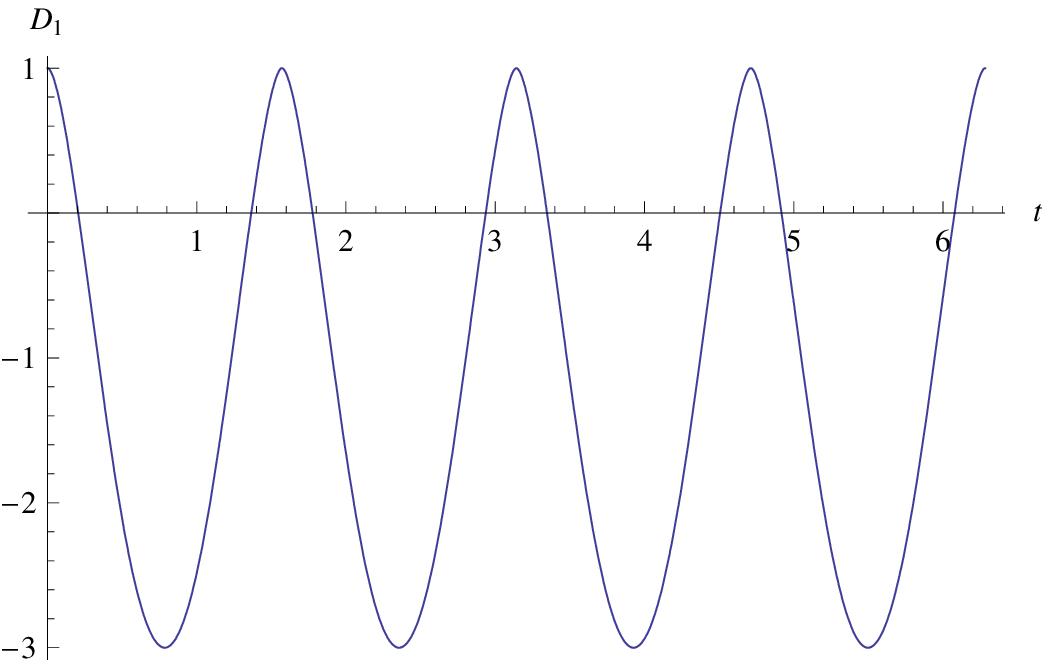}&
\includegraphics[height=4.0cm,width=5.5cm]{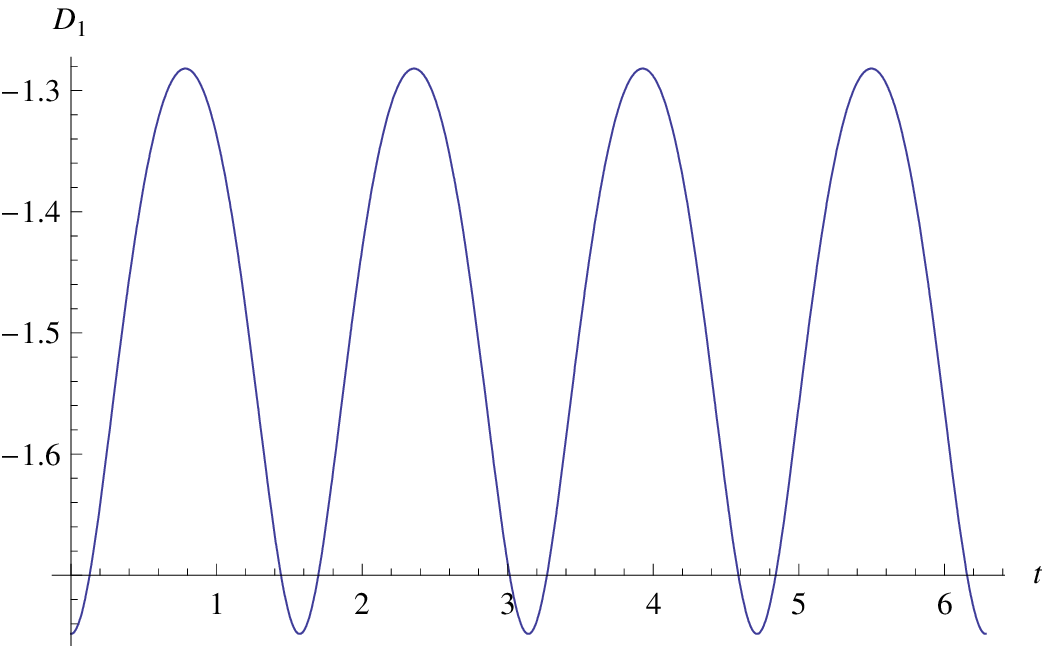}\\
(i)&(ii)
\end{array}
\]
\caption{$D_1$ for (i) three-qubit  pure GHZ state  and (ii) three-qubit pure W state .}
\end{figure}

\begin{figure}
\[
\begin{array}{cc}
\includegraphics[height=4.5cm,width=6.5cm]{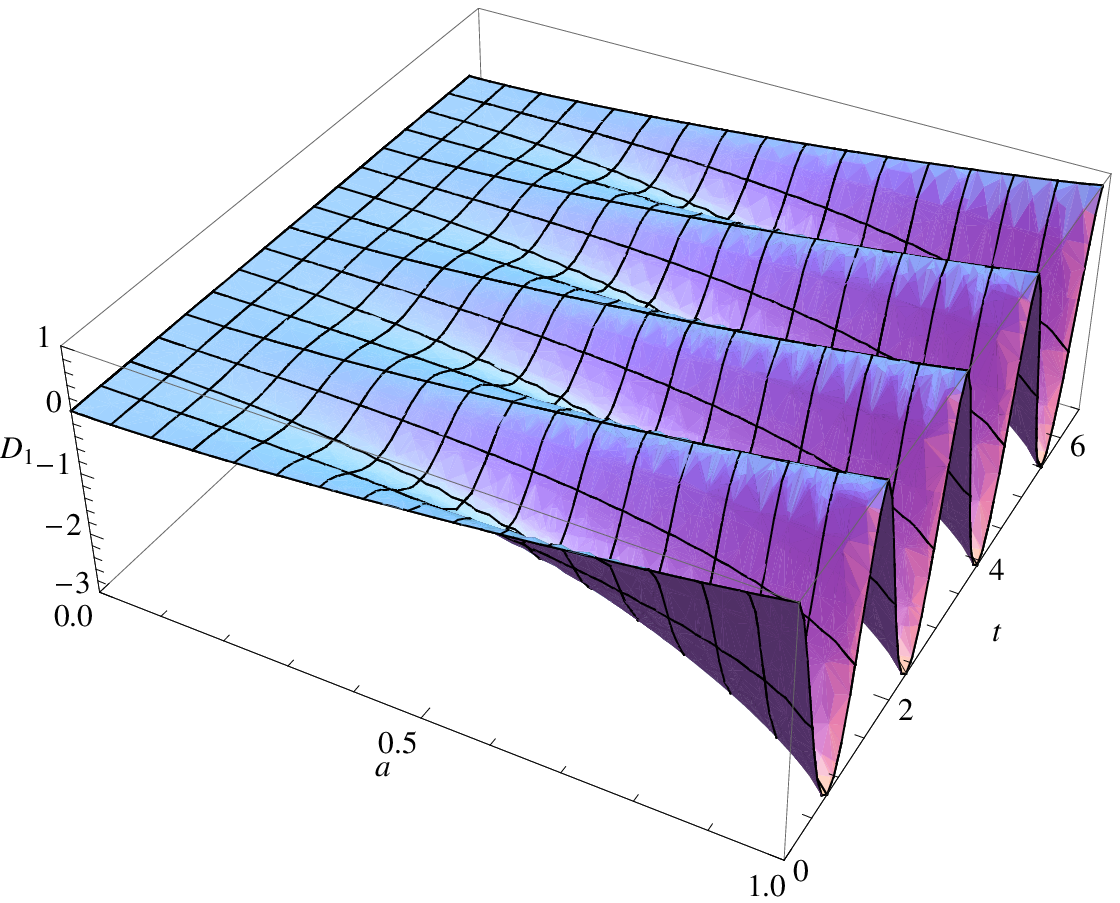}&
\includegraphics[height=4.5cm,width=6.5cm]{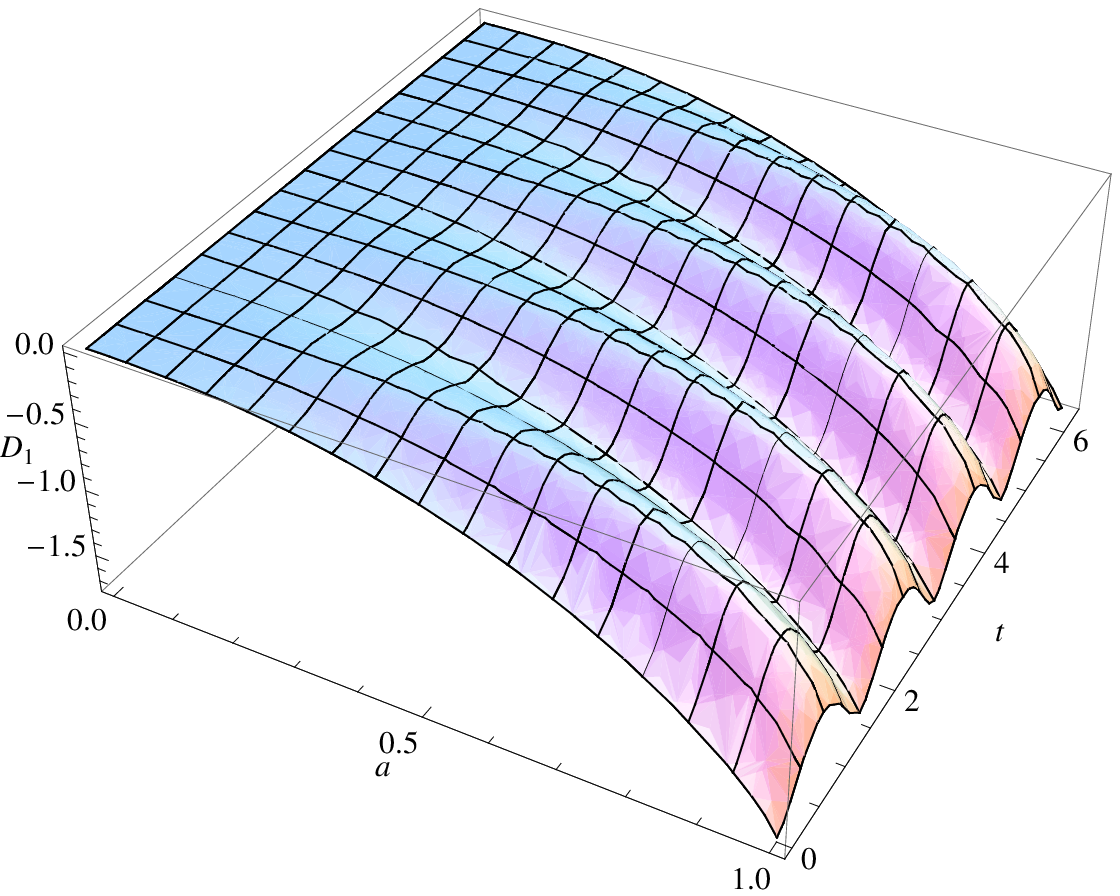}\\
(i)&(ii)\\
\includegraphics[height=4.5cm,width=6.5cm]{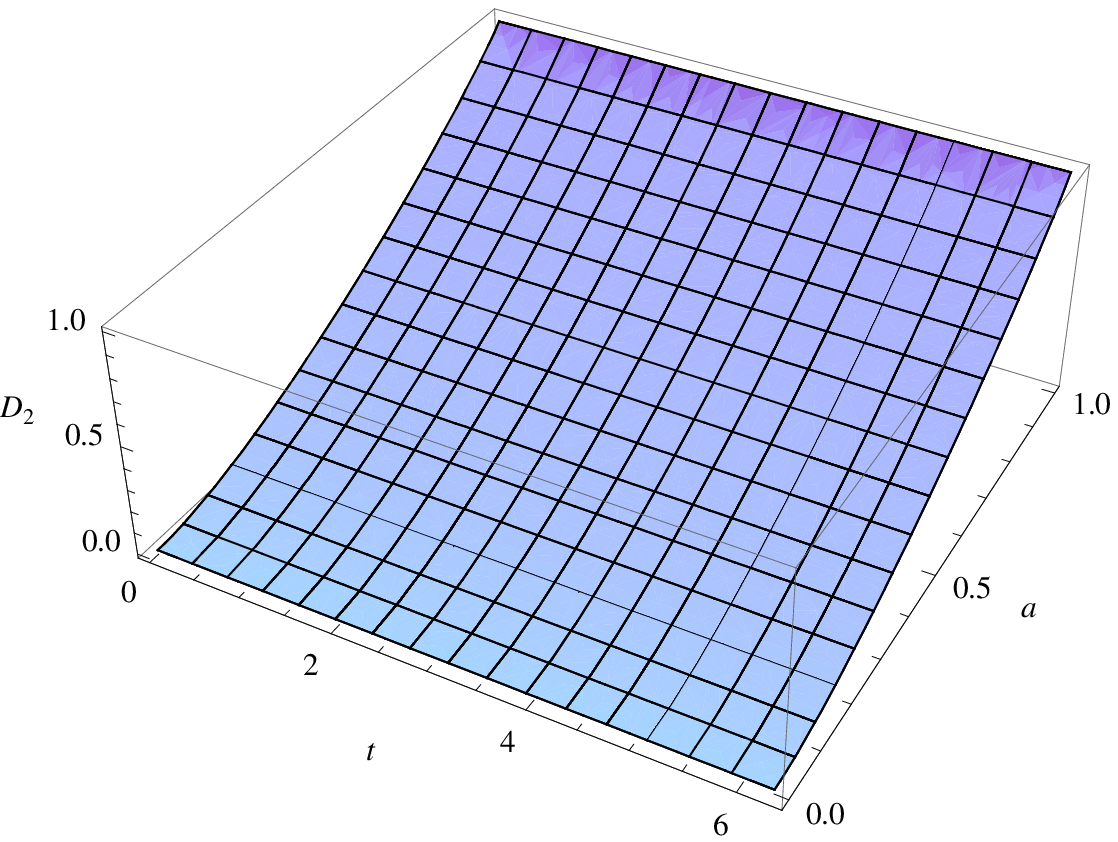}&
\includegraphics[height=4.5cm,width=6.5cm]{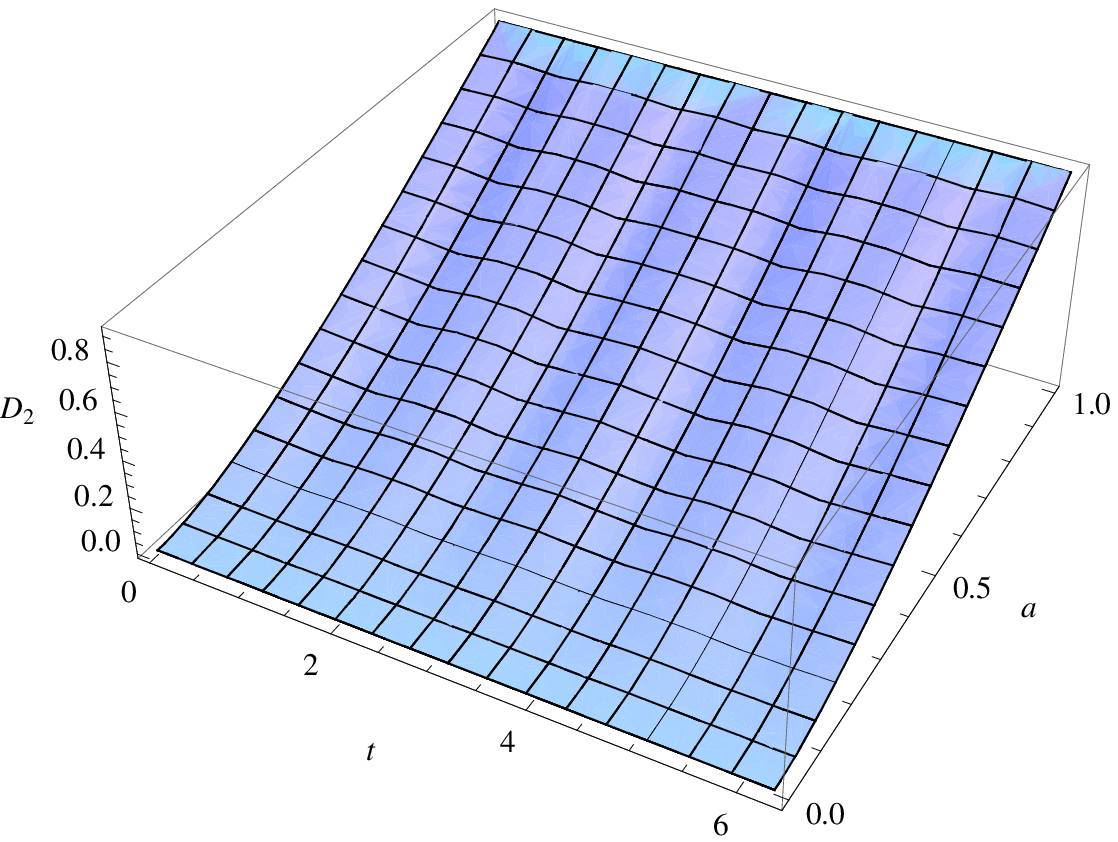}\\
(iii)&(iv)
\end{array}
\]
\caption{$D_1$ and $D_2$ for three-qubit  Mixed GHZ state (figures (i), (iii)) and three-qubit  Mixed W state (figures (ii), (iv)).}
\end{figure}
\end{center}
\end{widetext}

\end{document}